\documentclass[11pt,a4paper]{article}
\pdfoutput=1
\usepackage{jcappub}
\usepackage[utf8]{inputenc}
\usepackage{aas_macros}
\usepackage{amsfonts}
\usepackage{amsmath}
\usepackage{bm}
\usepackage{amssymb}
\usepackage{graphicx}
\usepackage[dvipsnames]{xcolor}
\usepackage{xfrac}
\usepackage{hyperref}
\usepackage{nameref}
\usepackage{cleveref}
\usepackage{multirow}
\usepackage{soul}
\usepackage{url}
\usepackage[thinlines]{easytable}
\crefformat{section}{\S#2#1#3}
\hypersetup{colorlinks=true,allcolors=teal}

\defcitealias{Arya_2023}{Paper I}
\defcitealias{Arya_2024a}{Paper II}

\providecommand{\keywords}[1]
{
  \small	
  \textbf{\textit{Keywords---}} #1
}

\newcommand{\Mpch}{\ensuremath{h^{-1}{\rm Mpc}}}
\newcommand{\cMpch}{\ensuremath{h^{-1}{\rm cMpc}}}

\newcommand{\avg}[1]{\ensuremath{\left\langle \,#1\, \right\rangle}}

\newcommand{\der}{\ensuremath{{\rm d}}}

\newcommand{\be}{\begin{equation}}
\newcommand{\ee}{\end{equation}}

\renewcommand{\citet}{\cite}

\title{\boldmath Covariance matrices for the Lyman-$\alpha$ forest using the lognormal approximation}

\author[a,1]{B. Arya,\note{Corresponding author.}}
\author[a]{A. Paranjape.}
\author[b]{T. Roy Choudhury,}

\affiliation[a]{Inter-University Centre for Astronomy \& Astrophysics,\\ Ganeshkhind, Post Bag 4, Pune 411007, India}
\affiliation[b]{National Centre for Radio Astrophysics, TIFR,\\Post Bag 3, Ganeshkhind, Pune 411007, India}

\emailAdd{bharya@iucaa.in}
\emailAdd{aseem@iucaa.in}
\emailAdd{tirth@ncra.tifr.res.in}

\abstract{
We investigate the nature of correlations in the small-scale flux statistics of the Lyman-$\alpha$ (Ly$\alpha$) forest across redshift bins. Understanding and characterising these correlations is important for unbiased cosmological and astrophysical parameter inference using the Ly$\alpha$ forest. We focus on the 1-dimensional flux power spectrum (FPS) and mean flux ($\bar F$) simulated using the semi-numerical lognormal model we developed in earlier work. The lognormal model can capture the effects of long wavelength modes with relative ease as compared to full smoothed particle hydrodynamical (SPH) simulations that are limited by box volume. For a single redshift bin of size $\Delta z\simeq 0.1$, we show that the lognormal model predicts positive cross-correlations between $k$-bins in the FPS, and a negative correlation for $\bar F\times$ FPS, in qualitative agreement with SPH simulations and theoretical expectations. For measurements across two neighbouring redshift bins of width $\Delta z$ each (using long flux skewers of length $2\Delta z$ that are `split' in half), the lognormal model predicts an \emph{anti-correlation} for FPS $\times$ FPS and a \emph{positive correlation} for $\bar F\times$ FPS, caused by modes with the longest wavelengths. This is in contrast to SPH simulations which predict a negligible magnitude for cross-redshift correlations derived from such `split' skewers, and we discuss possible reasons for this difference. Finally, we perform a preliminary test of the impact of neglecting long wavelength modes on parameter inference, finding that whereas the correlation structure of neighbouring redshift bins has relatively little impact, the absence of long wavelength modes in the \emph{model} can lead to $\gtrsim2-\sigma$ biases in the inference of astrophysical parameters. Our results motivate a more careful treatment of long wavelength modes in analyses that rely on the small-scale Ly$\alpha$ forest for parameter inference.
}


\begin{document}
\label{firstpage}
\maketitle
\flushbottom

\section{Introduction}

With the advent of SDSS eBOSS \citep{schelgel_2015, eboss_2016, blomqvist_2019, eboss_2021, nilipour_2022}, ongoing DESI \citep{ribera_2018, karacayli_2020, walther_2021, desi_2022, satya_2022} and upcoming WEAVE \citep{dalton_2012, weave_2022} surveys, it will be possible to observe $\gtrsim$ 1 million spectra from fainter, more numerous background quasars in large cosmological volumes $\gtrsim 1 \: (h^{-1}\,{\rm cGpc})^3$.
The interpretation and measurements of cosmological and astrophysical parameters from such a large number of observed spectra would require a careful modelling of the Lyman-$\alpha$ (Ly$\alpha$) forest in the cosmological simulations. 

This modelling involves not only predictions for the expected signal of some Ly$\alpha$ flux statistic such as the mean flux $\bar F$ along a skewer, the 1-dimensional flux power spectrum (FPS) that measures 2-point correlations of fluctuations of flux along the line of sight, or the flux probability density function (FPDF), but also the expected \emph{covariance} of these statistics for a given survey specification. The calculation of a covariance matrix is a critical ingredient in building a likelihood function, using which one would perform parameter inference. There have been several studies of the covariance matrix of FPS statistics to date \citep[e.g.,][]{kim+04,zhan+05,pd+13,irsic+13,irsic+17,boera+19}. While earlier work on the subject has typically focused on diagonal terms of the covariance \citep[e.g.,][]{kim+04}, more recent work that used higher quality observational data as well as larger volume, high resolution simulations have also investigated the off-diagonal covariance structure \citep[e.g.,][]{pd+13,irsic+13,irsic+17,boera+19}. In the present work we approach the issue of off-diagonal terms, especially the ones that correlate multiple redshift bins, using a semi-analytical approximation to Ly$\alpha$ flux statistics.

In our earlier work \citet[][henceforth, Paper I]{Arya_2023} and \citet[][henceforth, Paper II]{Arya_2024a}, we have used the lognormal approximation \citep{cj91,gh96,bd97,cps01,csp01,viel+02, mattarrese_2002, qiu_2006, Farr_2020} to model the statistics \{$\bar F$, FPS, FPDF\} at redshifts $z\sim2.5$. A comparison with the high-resolution smoothed particle hydrodynamics (SPH) simulation suite Sherwood \citep{bolton+17-sherwood} in \citetalias{Arya_2023} showed that the lognormal model provides a good description of these statistics at $z=2.5$ and can accurately recover the astrophysical parameters describing the IGM state at this redshift (see below for details of the model). In \citetalias{Arya_2024a}, however, we found that the model fails to accurately model the FPDF at redshifts $z\gtrsim2.6$. Following common practice \citep{Borde_2014, Nasir_2016, Villasenor_2023, Irsic_2023, cabayolgarcia2023neural}, we therefore discarded the FPDF measurements in deriving our constraints. In the present work too, we will focus on the statistics \{$\bar F$, FPS\} throughout the discussion.

In the entire analysis of \citetalias{Arya_2023} and \citetalias{Arya_2024a}, we considered only a single redshift bin at a time, having width $\Delta z\simeq0.04$ and centered at one of the values $z\in\{2.0,2.1,2.2,2.3,2.4,2.5,\\2.6,2.7\}$.\footnote{This range of redshifts was chosen so as to ensure the approximate validity of the assumption that the IGM causing the Ly$\alpha$ forest is in a state of low density and temperature and that photoionization equilibrium holds. The width of each bin was chosen to match the redshift path length spanned by the SPH box side of $40\,\cMpch$ that we focused on.} Consequently, each redshift bin was treated as an independent data set with its own covariance matrix, calculated as described later. Physically, this amounts to assuming that flux statistics measured in different redshift intervals are sourced by independent quasar sightlines or skewers. In reality, however, it is quite common for a \emph{single} skewer to span redshift intervals substantially broader than $\Delta z=0.04$, i.e., multiple bins as defined by our configuration \citep[see, e.g., fig.~1 of][]{Gaikwad_2021}, where a typical Ly$\alpha$ skewer can span upto $\Delta z \sim 0.5$ ($\sim$ 400 $\Mpch$) at $z \sim 2.5$. Since it is not desirable to assume constant values of the IGM thermal parameters for such broad bins, it becomes important to ask whether the possible correlations between flux statistics \emph{across} redshift bins induced by the skewers that span these bins are relevant for parameter inference. This issue has not received much attention in the literature \citep[although see][whose results we also comment on later]{irsic+17}. Our goal in the present work is to understand the magnitude and relevance of these cross-redshift correlations.

We emphasize that current SPH simulations are not capable of satisfactorily addressing this issue. This is because the evaluation of the relevant covariance matrices necessarily involves generating `long' skewers that properly account for long wavelength correlations along a backward light cone. SPH simulations such as Sherwood are limited by the volume of their (typically periodic) boxes, which therefore also limits the longest wavelengths probed by them. Semi-numerical approximations such as the lognormal model can, in principle, circumvent this difficulty by modelling arbitrarily long skewers limited only by the validity of the underlying assumptions, provided the evolution of the IGM thermal and photoionization state along the light cone is accounted for. 

In this work, we will use the lognormal model to assess the nature of cross-redshift correlations across one pair of neighbouring bins, which is the fundamental building block for understanding the full multi-bin correlation structure. We investigate the effects of long wavelength modes on these correlations and quantify the impact of these correlations on estimates of the astrophysical parameters defining the lognormal model.

The paper is as organised follows. In section~\ref{sec:sims}, we briefly describe the lognormal model and Sherwood simulations, along with the methodology for calculating covariance matrices. We present our results in section~\ref{sec:covariance} and conclude with a discussion in section~\ref{sec:conclude}.

\section{Simulations and Methods}
\label{sec:sims}
Here we briefly describe our lognormal implementation from \citetalias{Arya_2023} and \citetalias{Arya_2024a}, and the SPH simulation suite Sherwood \citep{bolton+17-sherwood}. We also describe the basic technique used for all our covariance matrix calculations, with subsequent variations being discussed in the following section. We kindly refer the reader to \citetalias{Arya_2023} and \citetalias{Arya_2024a} for further details of the basic techniques described in this section. 

\subsection{Lognormal approximation}
The lognormal model for the Ly$\alpha$ forest starts by assuming that the 3-dimensional \emph{linear} baryonic density field is described by the power spectrum
\begin{equation}
P_{\rm L,3D}(k,z) = D(z)^2\,P_{\rm L,dm}(k)\,{\rm e}^{-2x_{\rm J}(z)^2k^2}
\end{equation}
where $P_{\rm L,dm}(k)$ is the linear theory dark matter power spectrum extapolated to $z=0$, $D(z)$ is the linear theory growth factor normalised so that $D(0)=1$, and $x_{\rm J}(z)$ is the Jeans length which approximates the effects of pressure smoothing at small scales and is treated as a free parameter in the model. 

This power spectrum is averaged over two angular directions giving its 1-dimensional counterpart, which is used to generate a realisation of the \emph{1D linear} baryonic density field $\delta_{\rm b}^{\rm L}(x,z)$ and the corresponding velocity fluctuation $v_{\rm b}^{\rm L}(x,z)$, where $x$ labels the spatial direction along the line of sight and $z$ accounts for explicit time dependence. The realisation is drawn over a spatial distance $\Delta x$ corresponding to a redshift interval $\Delta z$, centered at the redshift $z$ under consideration, and has $\sum_x\delta_{\rm b}^{\rm L}(x,z)=0$ where the sum is over each individual sightline or skewer. Below, when comparing to SPH simulations we will use the choices $\Delta x=40\,\cMpch$ ($\Delta z\simeq 0.04$) and $\Delta x=80\,\cMpch$ ($\Delta z\simeq 0.08$). When discussing the broader structure of the covariance matrix in the lognormal model, we will also use other choices for $\Delta x$ and hence $\Delta z$.

Given the realisation of the linear field $\delta_{\rm b}^{\rm L}(x,z)$, the eponymous assumption of the model then says that the \emph{1-dimensional nonlinear} baryonic number density $n_{\rm b}(x,z)$ is
\begin{equation}
n_{\rm b}(x,z) \equiv \bar n_{\rm b}(z)\,\Delta_{\rm b}(x,z) = \bar n_{\rm b}(z)\,{\rm e}^{\delta_{\rm b}^{\rm L}(x,z)-\avg{(\delta_{\rm b}^{\rm L})^2}/2}\,,
\end{equation}
where $\bar n_{\rm b}(z)$ is the spatially averaged number density and $\avg{(\delta_{\rm b}^{\rm L})^2}$ is the variance of the linear 1-dimensional field (which is finite due to the Jeans smoothing). The corresponding nonlinear velocity field is assumed to be the same as the linear one, which is a safe assumption for standard hierarchical growth of structure \citep{paddy-strucform}.

The next step is the assumption of photoionization equilibrium, which is expected to hold at the redshifts $2\leq z\lesssim3$ that we focus on. One then calculates the neutral hydrogen density field $n_{\rm HI}$ which obeys
$n_{\rm HI} \propto n_{\rm b}(x,z)^2\,T(x,z)^{-0.7}/\Gamma_{12}(z)$,
where the proportionality constant depends on fundamental quantities, $\Gamma_{12}(z)$ is the photoionization rate in units of $10^{-12}\,{\rm s}^{-1}$ and $T(x,z)$ is the spatially varying temperature of the IGM, which is approximated (self-consistently with the assumption of photoionization equilibrium) as a power law in the density fluctuation: $T(x,z) = T_0(z)\,\Delta_{\rm b}^{\gamma(z)-1}$, where $T_0(z)$ is the temperature at mean density and $\gamma(z)$ is the equation of state coefficient. $\Gamma_{12}$, $T_0$ and $\gamma$ are three more free parameters in the model.

Finally, the optical depth $\tau(x,z)\sim\int\der l\,n_{\rm HI}\,\sigma$ with absorption cross-section $\sigma$ is calculated by accounting for redshift space distortions due to the velocity field $v_{\rm b}^{\rm L}$ and additionally accounting for thermal and natural broadening using the Voigt kernel. (The broadening kernel thus itself depends on the IGM temperature and hence the density fluctuations $\Delta_{\rm b}$, making this operation highly nonlinear). This gives us one realisation of the flux field $F={\rm e}^{-\tau}$, to which we add realistic noise as described in \citetalias{Arya_2023}.

We argued in \citetalias{Arya_2024a} that this model, with 4 free parameters $\{x_{\rm J},T_0,\gamma,\Gamma_{12}\}$ is unable to capture the redshift evolution of \{$\bar F$,FPS\} over the interval $2\leq z\leq 2.7$, primarily because the lognormal assumption does not capture the true non-Gaussianity of the baryonic density fluctuations $\Delta_{\rm b}$. To partially account for this, we introduced a scaling parameter $\nu$ for the linear field that sends $\delta_{\rm b}^{\rm L}\to\nu\,\delta_{\rm b}^{\rm L}$. The resulting 5-parameter model, which is still lognormal, has additional freedom to adjust the width of the density field $\delta_{\rm b}^{\rm L}$ and hence of $\Delta_{\rm b}$. In \citetalias{Arya_2024a} we showed that this leads to acceptable fits for \{$\bar F$, FPS\} over the entire redshift range $2\leq z\leq 2.7$ which correctly recover the true values of $T_0$ and $\gamma$, but significantly overestimate the photoionisation rate $\Gamma_{12}$ at $z\gtrsim2.2$. We also gave physical arguments explaining the origin of this overestimation. Notably, the best fitting model at each redshift had $\nu$ significantly smaller than unity.

\subsection{Sherwood simulations}
Our benchmark model is provided by the publicly available Sherwood simulations suite \citep{bolton+17-sherwood}. These were performed with a modified version of the cosmological SPH code P-Gadget-3, an extended version of publicly available GADGET-2 code \citep{Springel_2005}\footnote{\url{https://wwwmpa.mpa-garching.mpg.de/gadget/}}. The Sherwood suite consists of cosmological simulation boxes with volume ranging from $(10\,\cMpch)^3$ to $(160\,\cMpch)^3$, with particle numbers ranging from $2\times 512^3$ to $2 \times 2048^3$. These box sizes and mass resolutions are suitable for studying the small scale structures probed by Ly$\alpha$ forest. 

Similar to \citetalias{Arya_2023}, our default configuration for covariance studies in this work is a box of volume $(40\,\cMpch)^3$ containing $2 \times 2048^3$ particles. Later, we will also show results from a $(80\,\cMpch)^3$ box with $2 \times 2048^3$ particles.  For all these simulation boxes, we extract the baryonic density $\Delta_{\rm b}$, velocity $v_{\rm b}$ and temperature $T$ along skewers (the skewer configuration is described in the next section). Subsequently, we follow exactly the same steps as described for the lognormal approximation to calculate the neutral hydrogen number density $n_{\rm HI}$ followed by the optical depth $\tau$ and flux $F$, except that we use $\Delta_{\rm b}$, $v_{\rm b}$ and $T$ derived from the SPH. We add noise to the flux field thus generated identically to the lognormal case, and then compute the statistics \{$\bar F$, FPS\}. This gives an SPH data set that the lognormal one can be compared to.

\subsection{Covariance matrix calculation}
\label{subsec:cov_calc}
The calculation of $\bar F$, FPS and their covariance for both SPH and lognormal is similar to \citetalias{Arya_2023} (see their equations~12-16). In particular, the choice of Fourier space configuration for measuring the FPS -- and hence the total number of data points in the data vector at each redshift ($1$ $\bar F$ and $11$ FPS measurements for a total of $12$ data points) -- is identical to that in \citetalias{Arya_2024a}. 

For the {\bf SPH simulations}, we were limited to $5000$ skewers drawn randomly from the relevant simulation box. The flux statistics in this case are averaged over 100 skewers at a time, picked randomly from these 5000, leading to 50 `realisations' of the SPH data vector. As discussed in \citetalias{Arya_2024a}, the choice of including $100$ skewers per realisation leads to a total redshift path length comparable to that probed by existing surveys. Since the resulting 50 realisations are not enough for a precise estimate of a $12\times12$ covariance matrix, we additionally perform jackknife resampling as in \citetalias{Arya_2024a} to reduce the noise of the estimate. In particular, for each realisation we create 100 `leave-one-out' jackknife samples and calculate a jackknife covariance using these. The final SPH covariance matrix for a given redshift bin is the arithmetic mean of these 50 jackknife covariance matrices.

For the {\bf lognormal model}, as discussed in \citetalias{Arya_2024a}, we increase the redshift path length of each realisation to twice that of the SPH (i.e., we use 200 skewers per realisation). In principle, this number could be increased further, but we stick to our choice from \citetalias{Arya_2024a} where we were limited by the computation time of likelihood evaluation in the parameter inference (since each likelihood evaluation takes a finite amount of time to average over 200 skewers). Although we restrict the number of skewers when defining each realisation, we do exploit the freedom of generating an arbitrary number of lognormal sightlines by averaging over 200 realisations (instead of the 50 we were restricted to use in SPH). We have checked that this leads to a well-converged covariance matrix, obviating the need for jackknife resampling.\footnote{Nevertheless, in one of the variations we discuss in the next section, we will also show some results where the lognormal covariance is estimated using jackknife resampling.}

We emphasize that the averaging over 100 (200) skewers per realisation for the SPH (lognormal model) is the equivalent of averaging \{$\bar F$, FPS\} over these many \emph{independent} quasar sightlines having equal redshift path length. The reason to perform this average is simply to beat down the effects of noise and sample variance.

One issue that arises in the calculation of the covariance matrix is its possible dependence on the model parameters (or physical choices, in the case of SPH). While it is not possible to address this question for the SPH, since we cannot perform variations of the Sherwood suite, the lognormal model does allow us to calculate such parameter dependent covariance matrices.
We have therefore investigated the sensitivity of the covariance matrix in the lognormal approximation to simple variations in the model parameters. We find that $\sim10\%$ variations in the parameters lead to no significant change in the covariance matrix elements. In keeping with standard practice, we therefore show all results below for the same lognormal configuration used for covariance calculations in \citetalias{Arya_2023} and \citetalias{Arya_2024a}; we fix the parameters $\{T_0,\gamma,\Gamma_{12}\}$ to their input or inferred values from the SPH (i.e., the `true' values) and set $x_{\rm J}=0.12\,\cMpch$ and $\nu=1$.\footnote{Since the best fitting models in \citetalias{Arya_2024a} are more than $\sim10\%$ discrepant from this parameter choice, it is in principle interesting to ask whether the covariance matrices evaluated at the best fitting model lead to substantially different parameter inference. However, as argued in \citetalias{Arya_2024a}, the lognormal model itself needs further improvement in order to accurately recover the astrophysical parameters $\{T_0,\gamma,\Gamma_{12}\}$. We therefore leave a detailed investigation of the parameter dependence of the covariance matrix to future work.}

Having obtained a covariance matrix $C_{ij}$ for any data vector with entries labelled by $i,j=1,2,\ldots$, we can calculate the corresponding correlation matrix ${\rm Corr}(i,j)$ using
\begin{equation}
{\rm Corr}(i,j) = C_{ij}/\sqrt{C_{ii}\,C_{jj}}\,,
\end{equation}
so that the entries of ${\rm Corr}(i,j)$ take values in the range $(-1,1)$. \emph{Throughout, our discussion will focus on these normalised correlation matrices in order to emphasize the nature of correlations.} We find that the diagonal terms of the covariance matrices are similar between SPH and lognormal, so we do not discuss them further.  Of course, when calculating a likelihood for parameter inference, we use the appropriate covariance matrix $C_{ij}$ (see section~3.3 of \citetalias{Arya_2023}).

\section{Correlation structure of Ly$\alpha$ flux statistics}
\label{sec:covariance}
In this section, we present our main results on various aspects of the correlation structure of the \{$\bar F$, FPS\} measurements at different redshifts.

\subsection{Basic correlation structure of \{$\bar F$, FPS\} }
\noindent
\begin{figure*}
\centering
\includegraphics[width=\textwidth]{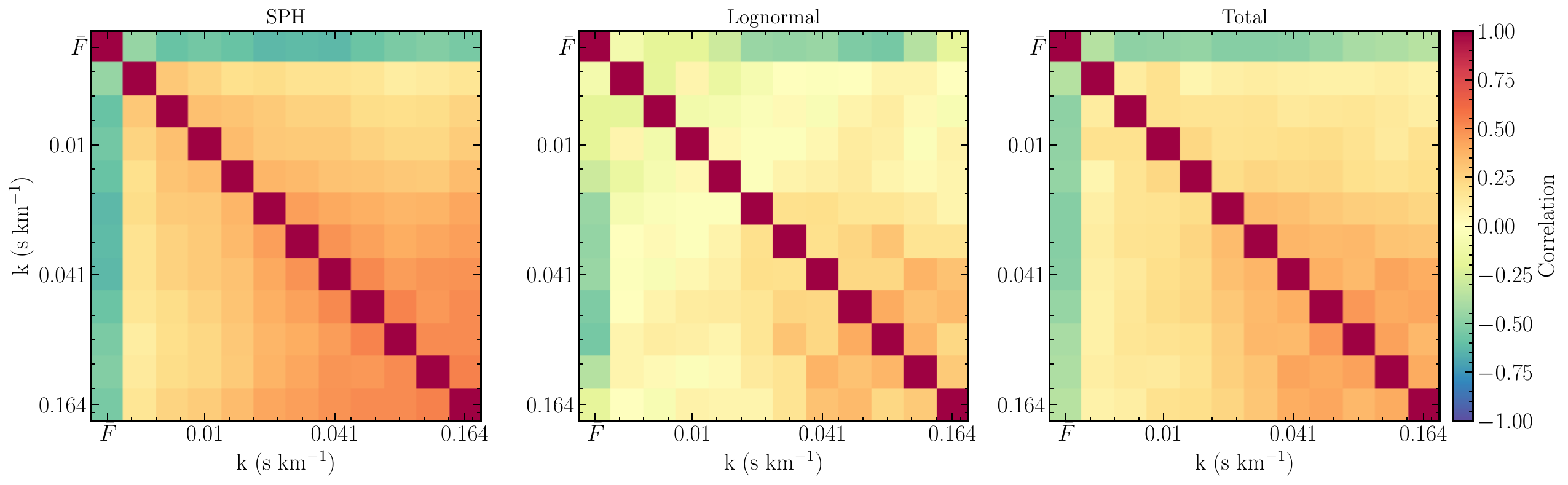}
\caption{Correlation matrix with cross-correlations between $\bar{F}$ and FPS for the SPH simulation \emph{(left panel)} and lognormal model \emph{(middle panel)} at $z=2.5$ for the skewer configurations discussed in the text. For reference, the \emph{right panel} shows the correlation matrix obtained from the sum of the SPH and lognormal covariance matrices (i.e., the covariance matrices were added and then normalised). 
}
\label{fig:corr_single}
\end{figure*}
We start by recalling the basic correlation structure of the \{$\bar F$, FPS\} statistics for a single redshift bin, which was already discussed in \citetalias{Arya_2023}. Fig.~\ref{fig:corr_single} shows the correlation matrices obtained at $z=2.5$ for the skewer configurations described in section~\ref{subsec:cov_calc} at the single redshift $z=2.5$. (This is similar to fig.~1 of \citetalias{Arya_2023}, except that we have discarded the FPDF measurements, and the skewer configuration used there was different.) 

\begin{figure*}
\centering
\includegraphics[width=0.75\textwidth]{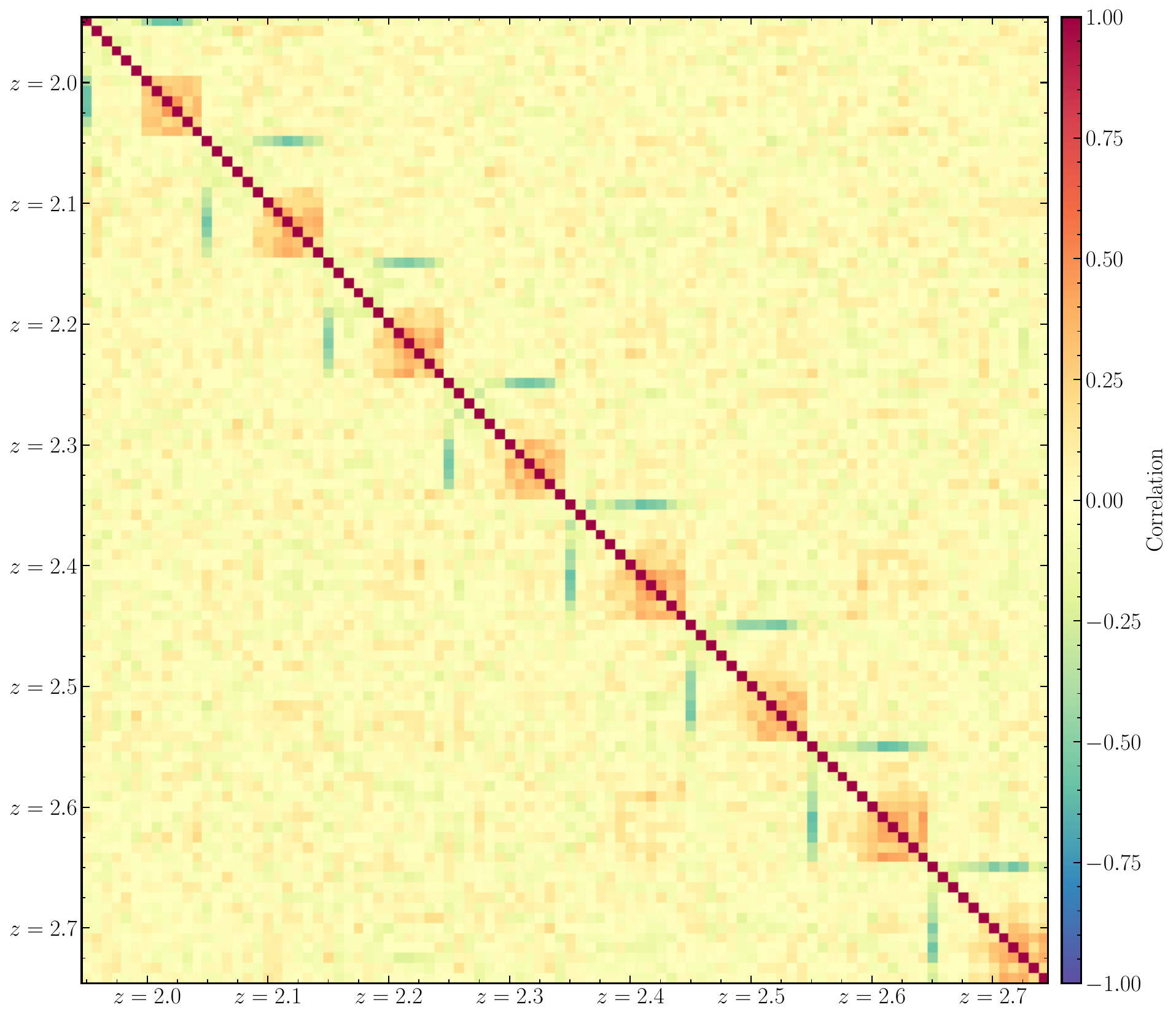}
\caption{Correlation matrix for \{$\bar F$, FPS\} statistics across redshifts $2.0\leq z\leq 2.7$ using the lognormal model, assuming that completely independent skewers contribute to each redshift. 
Cross-redshift correlations are negligible for this calculation by construction. The covariance was calculated using 200 realisations at each redshift, where each realisation was itself defined by averaging over 200 independent skewers with length $40\,\cMpch$ ($\Delta z\simeq0.04$).}
\label{fig:corr_combined}
\end{figure*}

We see noticeably large positive off-diagonal terms in the FPS $\times$ FPS correlation structure, and similarly large negative correlations in $\bar F\times$ FPS. The positive FPS $\times$ FPS correlations can be understood in terms of the highly nonlinear relation between the flux $F$ and the linear density fluctuation $\delta_{\rm b}^{\rm L}$. The negative correlation between $\bar F$ and FPS is worth some discussion. In the lognormal model, at large scales one can show that the measurements of $\bar F$ and FPS are nearly uncorrelated (see equation~4.2 of \citetalias{Arya_2024a}), since $\tau$ is approximately linear in $\delta_{\rm b}^{\rm L}$ at these scales. This is verified by the measured correlation matrix at $k\lesssim 0.02\,{\rm s\,km}^{-1}$. 
At larger $k$ (smaller length scales), the linear relation between $\tau$ and $\delta_{\rm b}^{\rm L}$ breaks down, and one must account for the fully nonlinear relation between $\delta_{\rm b}^{\rm L}$ and $\tau$. Since in the lognormal approximation $\tau \propto n_b^{\beta} \propto \mathrm{e}^{\beta \nu \delta_{\rm b}^{\rm L} - \beta \nu^2 \left\langle (\delta_{\rm b}^{\rm L})^2 \right \rangle / 2}$, with $\beta \approx 2.7 - 0.7 \gamma > 1$, we have
\begin{equation}
    \bar{\tau} \propto \exp\left[(\beta^2 - \beta) \nu^2 \left\langle (\delta_{\rm b}^{\rm L})^2 \right \rangle/2\right].
\end{equation}
If we now follow the argument presented in  \citetalias{Arya_2024a}, we find that to the lowest order $\bar{F} \propto \mathrm{e}^{-\bar{\tau}}$, and hence $\bar{F}$ along a specific skewer decreases when the variance of $\delta_{\rm b}^{\rm L}$ increases. On the other hand, a larger variance would lead to a higher amplitude of the FPS, which indicates that mean flux and FPS would be anti-correlated. At large scales $\tau \propto 1 + \beta \nu \delta_{\rm b}^{\rm L}$, and hence $\bar{\tau}$ does not depend on the variance in the lowest order leading to no correlation between $\bar{F}$ and FPS, consistent with the discussion in \citetalias{Arya_2024a}.

The correlation matrix for SPH appears smoother than the lognormal one, which is a consequence of the jackknife procedure used in the SPH. We expect that, if the lognormal covariance were averaged over a substantially larger number of realisations than the 200 used here, its correlation matrix would be similarly smooth.
More importantly, the magnitude of the correlations for any given low-$k$ bin is larger in SPH than in lognormal, and is more similar to the magnitude in the high-$k$ bins. This can be understood as arising from non-linearities at relatively larger scales (smaller $k$) that are correctly  captured by the SPH simulation but are missing in the lognormal approximation. 

Overall, though, the correlation structure seen in the lognormal is qualitatively similar enough to the SPH that we can safely use the lognormal covariance matrices for our subsequent discussion. As a simple reminder of the enforced lack of cross-redshift correlations in several current studies \citep[see, e.g.,][]{kim+04,pd+13,boera+19}, Fig.~\ref{fig:corr_combined} shows the correlation matrix obtained in the lognormal model over the full range of redshifts $2\leq z\leq 2.7$, using 8 independent sets of 200 realisations, implying different directions for skewers at each redshift. We clearly see that (a) the correlation structure seen in Fig.~\ref{fig:corr_single} is approximately repeated across the diagonal block (i.e., the redshift  evolution of the model parameters evidently does not significantly impact this structure) and (b) the cross-redshift terms are negligible by construction. 

It is worth noting that skewers extracted from SPH simulations from different redshifts are susceptible to being strongly correlated due to the finite volume of the box, especially if their spatial comoving location is not randomised. This effect is clearly visible in fig.~10 of \citet{irsic+17}; as noted by those authors, it arises due to their choice of allowing the same comoving skewer at neighbouring redshifts to approximate a light cone segment. The covariance derived from actual data by the same authors in their fig.~11 does not show the same spurious correlations, but does show some weak correlations that we comment on later.

\subsection{Effects of long wavelength modes}
\begin{figure*}
\centering
\includegraphics[width=0.9\textwidth]{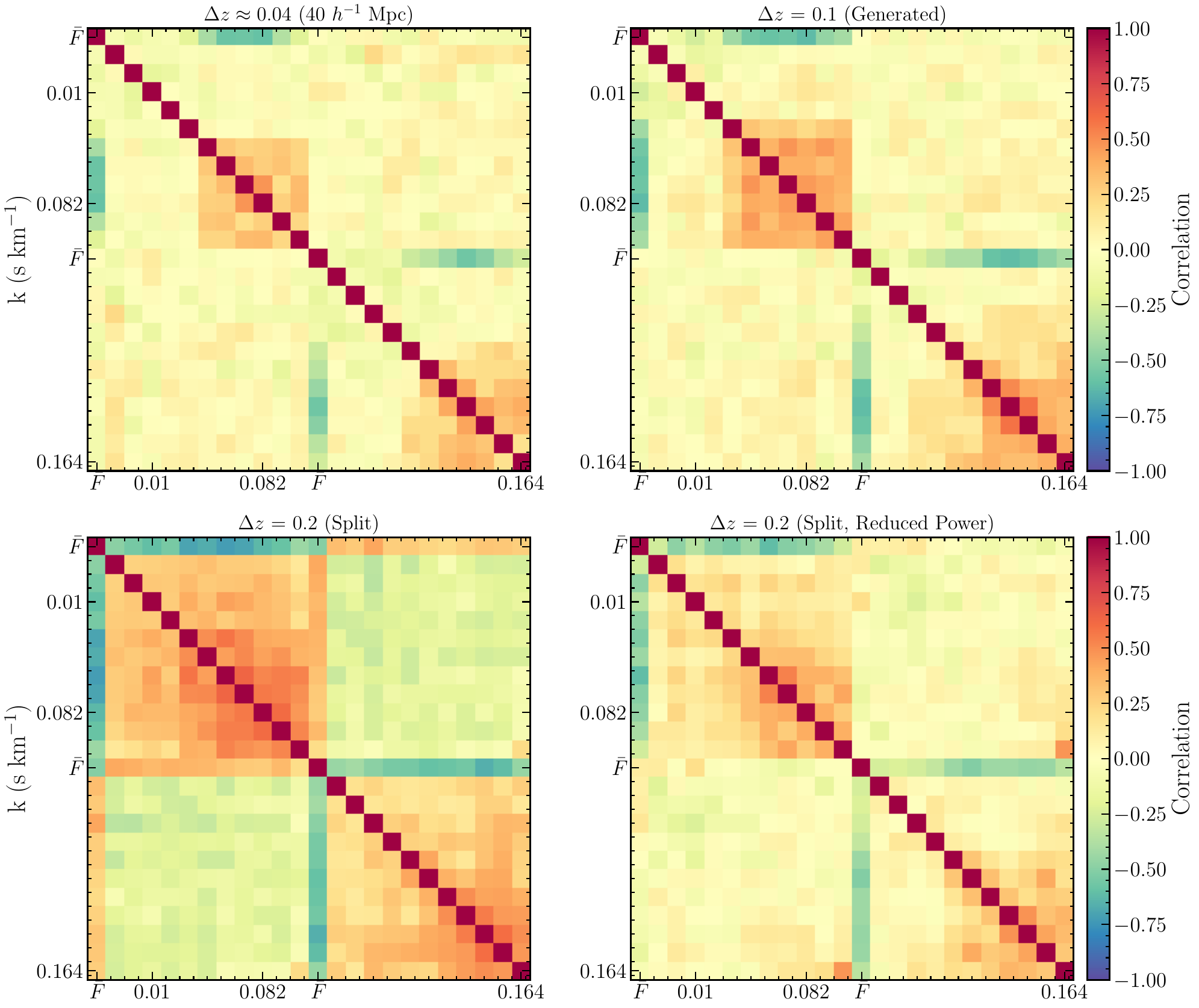}
\caption{Correlation matrices for two redshift bins centered at $z=2.0$ and $z=2.1$ (ordered as in Fig.~\ref{fig:corr_single}). \emph{(Top left panel):} Data for each bin was generated independently using $40\,\cMpch$ long skewers (having width $\Delta z\sim0.04$ each). These are the same as shown in the upper left $2\times2$ block of Fig.~\ref{fig:corr_single}. \emph{(Top right panel:)} Similar to the \emph{top left panel} but with skewers spanning the full bin interval $\Delta z = 0.1$ ($\sim 100\,\cMpch$ in length), still generated independently for each bin. \emph{(Bottom left panel:)} Correlation matrix obtained by generating skewers spanning redshift interval $\Delta z=0.2$ (length $\sim200\,\cMpch$) centered at $z=2.05$ and splitting each skewer in two halves of length $\Delta z=0.1$. \emph{(Bottom right panel:)} Same as the \emph{bottom left panel}, except that the power in two smallest $k$-modes in the 1D baryonic power spectrum is set to zero before generating the skewers. In each case, the covariance was calculated using $200$ realisations, each constructed by averaging the flux statistics over $200$ skewers. See text for a discussion.}
\label{fig:corr_cross_lgn_skewers}
\end{figure*}

\begin{figure*}
\centering
\includegraphics[width=0.9\textwidth]{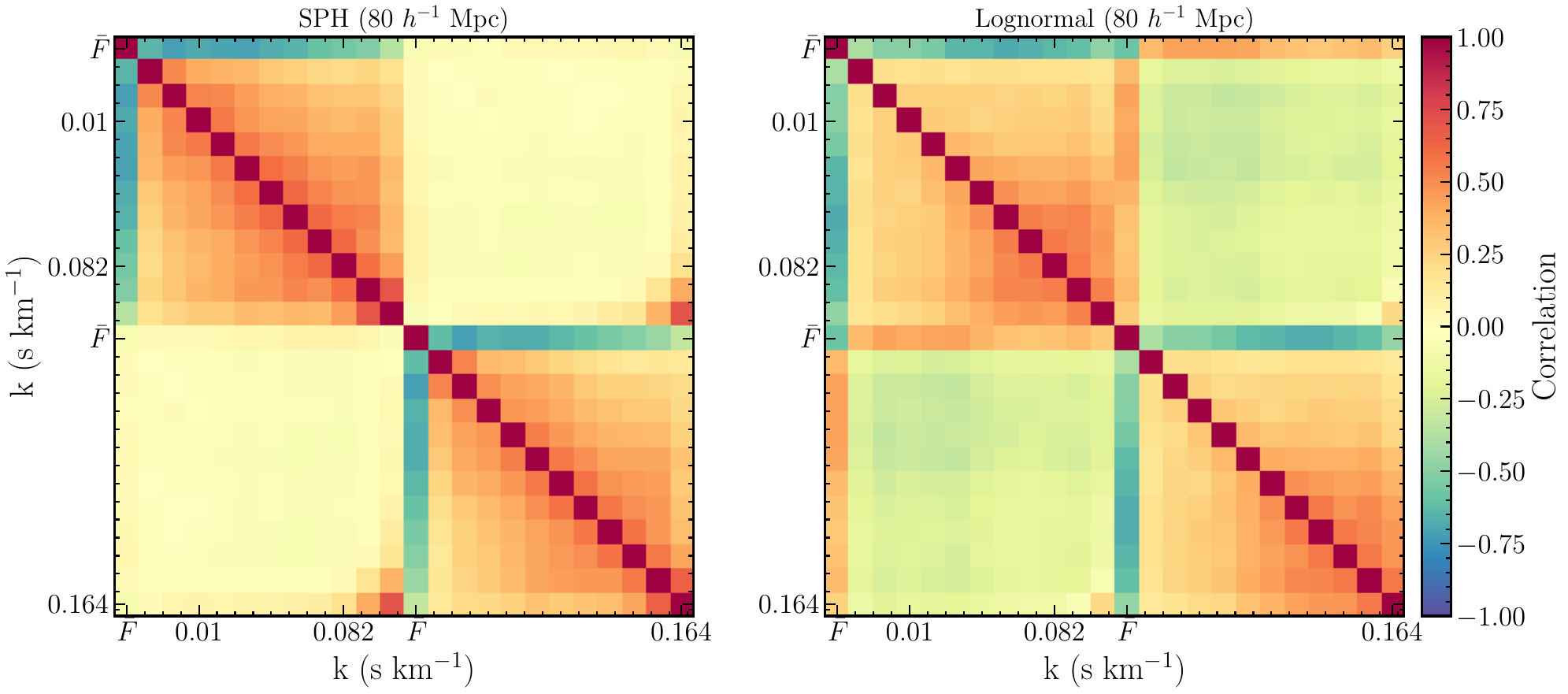}
\caption{Similar to the \emph{bottom left panel} of Fig.~\ref{fig:corr_cross_lgn_skewers}, now for an $80\,\cMpch$ configuration split in two equal halves, calculated using $5000$ independent skewers split into $100$ realisations averaging the flux statistics over $50$ skewers each and jackknife resampling the result. The \emph{left (right) panel} shows results using the SPH simulation (lognormal model). See text for further details and a discussion.}
\label{fig:corr_cross_lgn_sph}
\end{figure*}

\noindent
As mentioned in the introduction, in realistic data sets we expect any given quasar sightline to span multiple redshift bins. Since not all sightlines will span the same bins \citep[see fig.~1 of][]{Gaikwad_2021}, the resulting induced correlations across redshift bins can in principle be a complicated function of survey configuration. In this section, we simplify this discussion using some idealised configurations to better understand the basic correlations induced by long wavelength modes across multiple redshift bins. We build up our results in several steps in Figs.~\ref{fig:corr_cross_lgn_skewers} and~\ref{fig:corr_cross_lgn_sph}, each of which explores correlations between two neighbouring redshift bins centered at $z=2.0$ and $z=2.1$ as discussed next. 

As a reference, in the \emph{top left panel} of Fig.~\ref{fig:corr_cross_lgn_skewers} we repeat the correlation structure at $z=2.0$ and $z=2.1$ from Fig.~\ref{fig:corr_combined}; i.e., this panel shows the correlations for skewers generated independently  at each redshift, spanning a comoving length of $40\,\cMpch$ or a redshift interval of $\Delta z\simeq0.04$ around each redshift center, i.e., in the redshift ranges $1.98<z<2.02$ and $2.08<z<2.12$, respectively. As mentioned earlier, the cross-redshift contribution is negligible by construction. One issue with this configuration is that the skewer length does not span the entire redshift bin, meaning that the skewer is missing a range of available long wavelength modes. This is, of course, very relevant for covariance matrices estimated from SPH simulations whose boxes typically have relatively small lengths compared to the expected redshift intervals spanned by real data. To assess the effects of these missing modes in a \emph{single} bin (while still ignoring cross-redshift correlations), the \emph{top right panel} of Fig.~\ref{fig:corr_cross_lgn_skewers} shows the correlation structure obtained from skewers spanning $\Delta z=0.1$, i.e., in the ranges $1.95<z<2.05$ and $2.05<z<2.15$, respectively ($\sim2.5$ times longer than the ones in the \emph{top left panel}), but keeping all other details of the covariance calculation the same. We immediately see that the strength of correlations along the diagonal block has increased in general, with cross-$k$ bin correlations extending to relatively smaller $k$. In other words, the $40\,\cMpch$ configuration of the \emph{top left panel}, and possibly also the $\sim100\,\cMpch$ configuration of the \emph{top right panel}, is not fully converged with respect to the effects of long wavelength modes. And, as expected, the cross-redshift correlations in the \emph{top right panel} continue to be negligibly small.

Next, we consider the effects of modes with long enough wavelengths to couple the two bins. To do this we generated $200$ realisations of skewers spanning the redshift interval $\Delta z=0.2$ centered at $z=2.05$, i.e., the single redshift range $1.95<z<2.15$. In this case, we linearly interpolated the parameters $\{T_0,\gamma,\Gamma_{12}\}$ to their values at $z=2.05$ from the values at $z=2.0$ and $z=2.1$ that were used previously. We then `split' each of these long skewers into two equal halves, spanning the intervals $1.95<z<2.05$ and $2.05<z<2.15$, respectively. Each set of `split' skewers was then treated identically to the separate redshift skewers as before and the covariance matrix was calculated. The resulting correlation matrix is shown in the \emph{bottom left panel} of Fig.~\ref{fig:corr_cross_lgn_skewers}. Firstly, we see a substantially larger magnitude of auto-correlation (both FPS $\times$ FPS and $\bar F\times$ FPS) extending to smaller $k$ than in the \emph{top panels}, consistent with the argument that those measurements were not converged with respect to the presence of long wavelength modes. More interestingly, the cross-redshift terms now show a \emph{significant anti-correlation} between $k$-bins, and a \emph{significant positive correlation} between $\bar F$ and FPS. This may be compared with fig.~11 of \citet{irsic+17} which shows a bootstrap covariance matrix from observational data. While that estimate is noisy, it does support the presence of non-zero cross-redshift correlations in neighbouring bins, although the sign of the cross-correlation appears to depend on redshift. (Note that the data used by \citealp{irsic+17} spans the redshift range $3\leq z\leq4.2$ and their bin widths are $\Delta z=0.2$).

Clearly, the presence of long wavelength modes that span both redshift intervals has dramatically affected the cross-redshift correlations. To verify this, we perform the following test. We again generate 200 realisations of skewers spanning $\Delta z=0.2$ each, except that we set the power in the two smallest $k$ modes of the 1D baryonic power spectrum to zero before generating the sample. Note that the mean density fluctuation over each individual skewer, i.e. the skewer's zero mode, is zero by construction (we comment further on this below); we now additionally suppress two more long wavelength modes. The result is in the \emph{bottom right panel} of Fig.~\ref{fig:corr_cross_lgn_skewers} and clearly shows that the magnitude of auto-correlations and especially cross-redshift correlations is substantially suppressed. In fact, the result is closer to the \emph{top right panel} which used completely independent skewers in the two halves of the redshift interval.

Thus, non-zero initial long-wavelength power can generate significant cross-redshift correlations across redshift bins of width $\Delta z\simeq0.1$. It is also interesting to ask whether a non-zero value of the zero mode of the initial baryonic density (i.e., a scatter across skewers in the mean density calculated along each skewer) might also create an impact. This is not, in general, an easy problem to address. In the lognormal approximation, a reasonable approach would involve generating extremely long skewers (of comoving length $\gtrsim 1\,h^{-1}$Gpc, say), and then consider smaller subsets of redshift intervals $\Delta z\simeq0.1$ (or comoving length $\sim100\,\Mpch$ at $z\simeq2$). Due to the considerable time evolution along a backward light cone over Gpc lengths, one would have to properly account for various time dependent factors such as the $(1+z)^3$ dependence of the mean baryonic density, the time dependence of the growth factor, etc. Our current implementation of the lognormal model (which works at a single mean redshift), is not set up to do this correctly. An SPH simulation would also struggle to correctly account for this effect, due to its necessarily finite volume. For while it is true that each skewer drawn from an SPH box is linked in a complex, nonlinear and time-dependent manner to the initial data -- and is therefore not constrained by the requirement that the initial 1D baryonic field have a zero mean along that skewer -- it is not guaranteed that the dependence of this `effective zero mode' on modes with wavelength longer than the box size has been correctly captured.\footnote{
One might argue that the anti-correlation seen in the lognormal covariance is a trivial artefact of setting the zero mode (i.e., mean density) along each skewer to zero, which naturally anti-correlates the density in the two half-skewers. While this is correct for the \emph{zero mode} of these half-skewers, it is not obvious that the anti-correlation should trivially extend to higher $k$ modes as well.} 
We have seen above in the lognormal approximation (where we can perform these experiments), that the results of `boxes' of length even as large as $100\,\cMpch$ are not converged with respect to the inclusion of long wavelength power.

It is therefore not straightforward to assess the importance of the box-scale zero mode (or super-sample covariance). We are presently working on a variation of the lognormal approximation that will correctly account for various redshift dependent factors, along with the self-consistent inclusion of redshift dependent astrophysical parameters. In the meantime, as a toy exercise we consider the following comparison. We focus on $80\,\cMpch$ `boxes' at $z=2$ ($\Delta z\simeq 0.08$), which are readily accessible in both SPH and lognormal cases. In each case, we split the skewers into two halves of $\Delta z\simeq 0.04$ and calculate a covariance matrix as in Fig.~\ref{fig:corr_cross_lgn_skewers}. As in section~\ref{subsec:cov_calc}, we average over 100 independent skewers representing independent quasar sightlines to create one `realisation', and calculate the covariance matrix by averaging over many such realisations. Since the maximum number of skewers available to us in SPH is 5000 (which gives us 50 realisations), we ensure an apples-to-apples comparison by using 5000 skewers combined into 50 realisations for both SPH and lognormal. To reduce sampling noise, we further apply jackknife resampling as described in section~\ref{subsec:cov_calc} to both SPH and lognormal covariances.

The resulting correlation matrices are shown in Fig.~\ref{fig:corr_cross_lgn_sph}. We see that the lognormal correlation structure looks very similar to the \emph{bottom left panel} of Fig.~\ref{fig:corr_cross_lgn_skewers}, but the SPH structure shows negligible cross-redshift correlations. The auto-correlation structures in each case are quite similar, although the SPH shows enhanced off-diagonal terms as commented on earlier (see Fig.~\ref{fig:corr_single}). In other words, the box-scale zero mode effectively randomises the cross-redshift terms in SPH, while it is absent in the lognormal. Since our earlier tests suggest that long wavelength modes tend to create correlated rather than uncorrelated scatter, for the time being we will assume that the lognormal correlation structure of Figs.~\ref{fig:corr_cross_lgn_skewers} and~\ref{fig:corr_cross_lgn_sph} is closer to the truth, and return to a more detailed comparison in future work.

\subsection{Effects on parameter estimation}
\noindent
We now investigate the potential impact of the presence of cross-redshift correlations as revealed by the lognormal approximation on parameter estimates. Since our focus is on understanding the impact of different correlation structures, we do not use SPH data in this exercise so as to avoid any systematic biases induced by the comparison of the lognormal model to SPH data (\citetalias{Arya_2024a}). Rather, we use the lognormal model itself to generate a mock dataset as described in section~\ref{subsec:cov_calc} which we then analyse using two different covariance configurations.

The mock dataset is created from 40000 skewers of size $80\,\cMpch$ at $z = 2$ using the same parameter value choices as used above. Each skewer is split in two equal halves and the flux statistics \{$\bar F$, FPS\} are calculated separately for each half. We average the flux statistics over 200 skewers at a time to create 200 realizations of the mock data. (This is similar to the configuration shown in the \emph{middle panel} of Fig.~\ref{fig:corr_single}, except for the increased length of each skewer which is then split in two.) We use one of the realisations as the `data', and use all 200 to calculate the full covariance matrix.\footnote{This covariance matrix is different from the one shown in the \emph{right panel} of Fig.~\ref{fig:corr_cross_lgn_sph} which is calculated from 5000 skewers split into 100 realizations followed by jackknife resampling in order to be compared to SPH version.}
We set the diagonal error on $\bar F$ to the value obtained from taking the variance over all lognormal realisations. This is different from our choice in \citetalias{Arya_2024a} where this error was set to $5\%$ of $\bar F$ in the input model. Our choice gives higher weight to $\bar F$ than in \citetalias{Arya_2024a}, and we comment on this below.

\begin{figure*}
\centering
\includegraphics[width=0.415\linewidth]{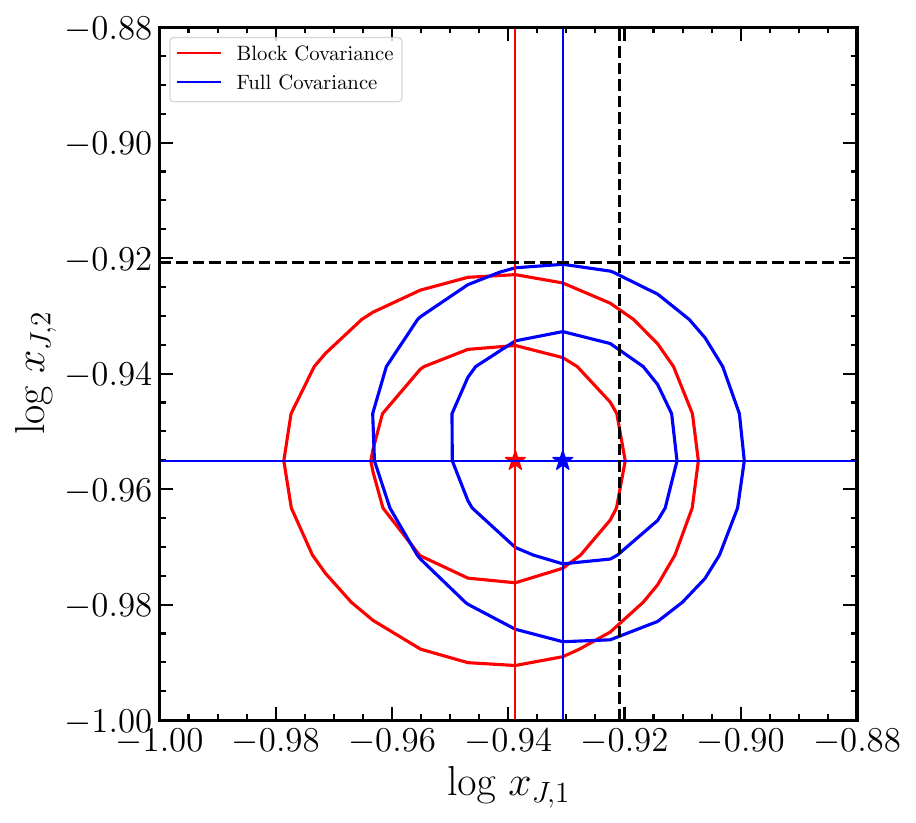}
\includegraphics[width=0.4\linewidth]{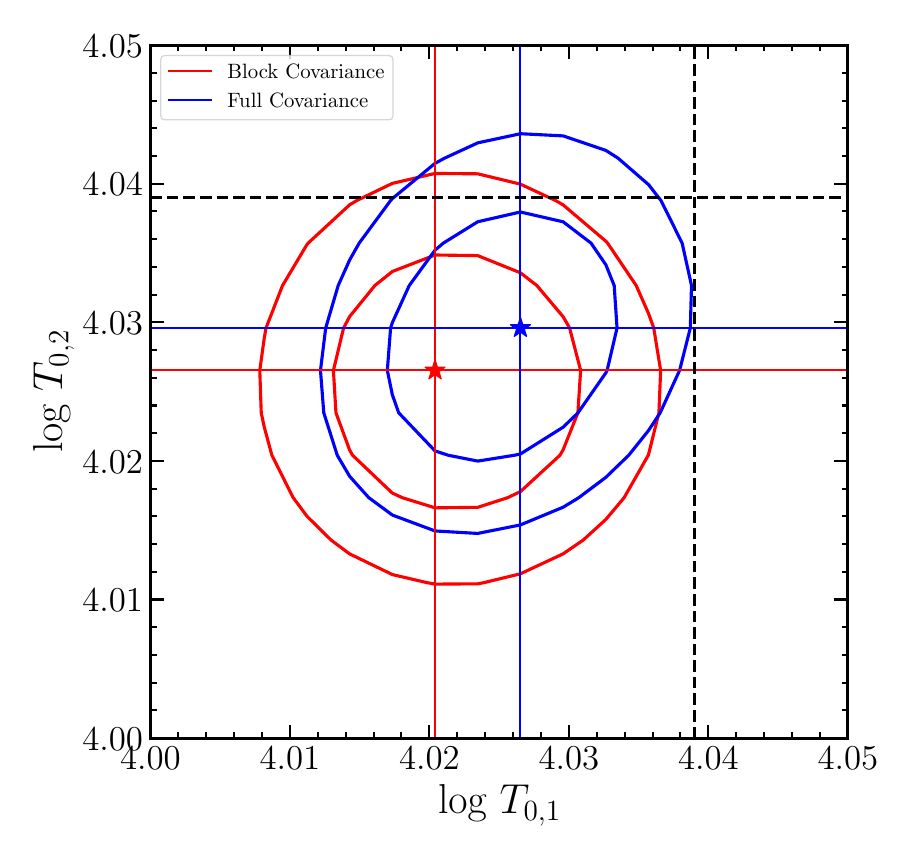}
\label{fig:xJ1xJ2_T1T2}
\caption{Parameter constraints on the `redshift split' versions of $\log x_{\rm J}$ \emph{(left panel)} and $\log T_0$ \emph{(right panel)} using the full covariance including cross-redshift correlations (blue curves and stars) and its block diagonal counterpart (red curves and stars). The inner (outer) curves show $68.3\%$ ($95.4\%$) confidence regions, with the best fit marked by the stars and horizontal and vertical coloured lines. Horizontal and vertical dashed black lines show the input (`true') values of each parameter. See text for a discussion.}
\label{fig:xJ_T0}
\end{figure*}

\begin{figure*}
\centering
\includegraphics[width=0.31\textwidth]{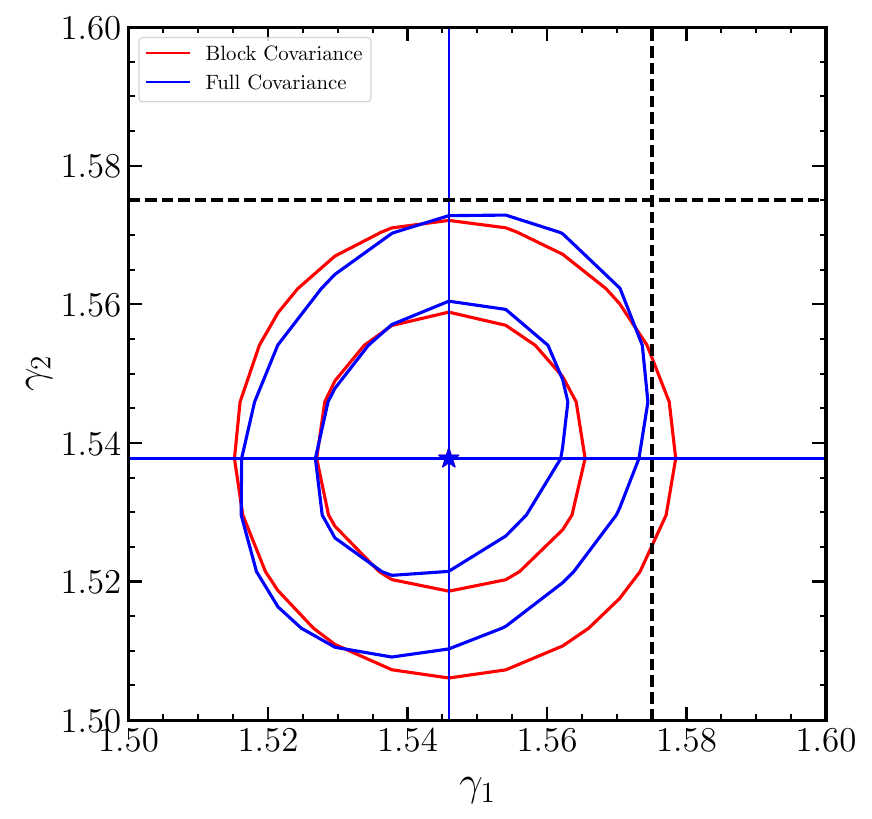}
\includegraphics[width=0.32\textwidth]{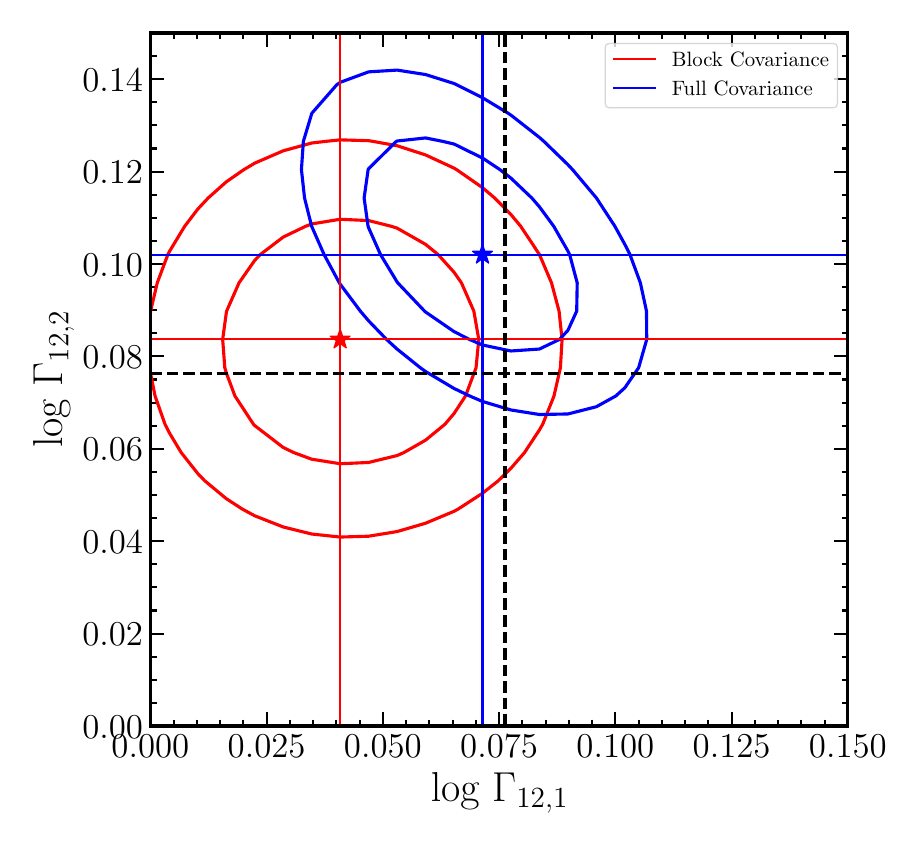}
\includegraphics[width=0.31\textwidth]{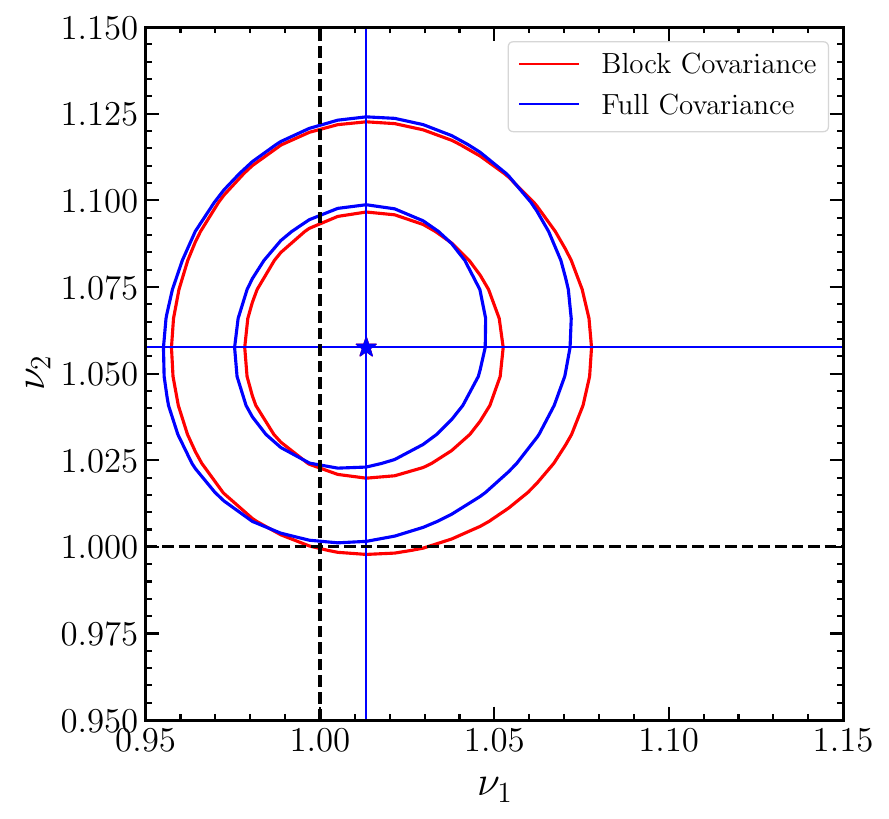}
\caption{Same as Fig.~\ref{fig:xJ_T0}, for the parameters $\gamma$ \emph{(left panel)}, $\log\Gamma_{12}$ \emph{(middle panel)} and $\nu$ \emph{(right panel)}. See text for a discussion.}
\label{fig:g1g2_G1G2_nu1nu2}
\end{figure*}

We analyse this data by fixing all but one parameter (say $x_{\rm J}$) to the input values, and allow $x_{\rm J}$ to separately vary in each of the two $40\,\cMpch$ halves of every skewer. We call these two variables as $x_{\rm J,1}$ and $x_{\rm J,2}$, respectively. The model thus uses skewers of only half the total length (i.e., $40\,\cMpch$) with the same resolution and combines two such \emph{independently generated} skewers with parameters $x_{\rm J,1}$ and $x_{\rm J,2}$, respectively, to describe a single long data skewer of length $80\,\cMpch$. Ideally, the model itself should account for long wavelength correlations by generating skewers that are long enough to describe the given data along with accounting for all possible long wavelength modes. As argued above, this is presently unfeasible, so we restrict ourselves to this simpler toy exercise. We then construct a Gaussian likelihood ($\propto{\rm e}^{-\chi^2/2}$), where the $\chi^2$ is defined either with the full covariance matrix spanning $80\,\cMpch$ and accounting for cross-redshift correlations, or a block diagonal covariance that sets the cross-correlations between the two half-skewers (i.e., `cross-redshift' terms) to zero by hand. We show the results of the $\chi^2$ minimisation exercise in these two cases when treating $x_{\rm J,1}$ and $x_{\rm J,2}$ as free parameters using blue and red contours for the full and block diagonal covariances, respectively, in the \emph{left panel} of Fig.~\ref{fig:xJ_T0}. The \emph{right panel} of Fig.~\ref{fig:xJ_T0} and the three panels of Fig.~\ref{fig:g1g2_G1G2_nu1nu2} similarly show results when treating $T_0$, $\gamma$, $\Gamma_{12}$ and $\nu$ as free parameters in the two half-skewers, respectively.

We see that, overall, the inclusion of cross-redshift correlations has a relatively minor effect on all parameters, as compared to the analysis with a block diagonal covariance that neglects these effects. The exception is $\Gamma_{12}$ (\emph{middle panel} of Fig.~\ref{fig:g1g2_G1G2_nu1nu2}) for which the values in the two halves become substantially anti-correlated when accounting for the full covariance structure; however, the constraints are still statistically consistent with those obtained from the block diagonal covariance analysis. Interestingly, for every parameter \emph{except} $\Gamma_{12}$, both analyses (with or without cross-redshift terms) exclude the input values of the respective parameters at $\gtrsim 95\%$ confidence. This is possibly due to the error we make in constructing the \emph{model} skewers, which are $40\,\cMpch$ in length and independently drawn to describe the two halves of the $80\,\cMpch$ data skewer, and thus do not correctly account for the presence of long wavelength modes in the `data'. Thus, at the least our lognormal model should be modified to exclude the stringent requirement of skewer-wise fixed mean density; the scatter in this mean density also must be generated by correctly accounting for its correlation with long wavelength modes. 

Finally, we also repeated the analysis using \emph{only} the FPS measurements in calculating the likelihood, which is closer to the analysis of \citetalias{Arya_2024a} mentioned above. We found that, for all parameters, the constraints were broader to some degree or other than those in Figs.~\ref{fig:xJ_T0} and~\ref{fig:g1g2_G1G2_nu1nu2}, making the discrepancy with the input values less pronounced. Apart from this, the constraints on $x_{\rm J}$, $T_0$, $\gamma$ and $\Gamma_{12}$ are very similar to those shown here, while those on $\nu$ are broad enough to make them fully consistent with the input values at the $2\sigma$ level. The constraints on $\nu$ with this reduced weight for $\bar F$ also show larger differences between the analyses with and without cross-redshift correlations (while still keeping these analyses statistically consistent with each other). Given the various systematic uncertainties in our overall analysis, however, we have chosen not to pursue these differences until we improve our model further.

\section{Conclusion}
\label{sec:conclude}
The covariance matrix of Ly$\alpha$ flux statistics is a crucial ingredient in likelihood analyses for parameter inference. We have investigated the structure of the correlations between the mean flux ($\bar F$) and flux power spectrum (FPS) using the lognormal approximation at redshifts $z\gtrsim2$.

Our main aim was to assess the importance of long wavelength modes in fixing the correlation structure of the \{$\bar F$, FPS\} statistics, especially `cross-redshift' correlations that link neighbouring redshift bins. As we argued, such correlations are expected to arise in realistic data sets due to individual quasar sightlines (skewers) that extend over distances larger than the choice of width of individual bins (which is typically $\Delta z\simeq0.1$ in observational studies). We also argued that the lognormal approximation provides us with a unique opportunity to study such correlations, since it allows us to generate skewers of arbitrary length, restricted only by the validity of the various assumptions underlying the model (photoionization equilibrium, optical thinness, absence of shocks in the IGM, slow evolution of astrophysical parameters, etc). We focused on the redshift range $2.0\leq z\leq 2.2$ where these assumptions are expected to be valid.

We demonstrated that the presence of long wavelength modes leads to an \emph{anti-correlation} between FPS measurements, and a corresponding \emph{positive correlation} between $\bar F$ and FPS measurements, in neighbouring redshifts (Fig.~\ref{fig:corr_cross_lgn_skewers}). Although this cross-redshift correlation structure -- which we showed arises from the 2-3 longest wavelength modes available in the grid used to generate the lognormal skewer -- is not seen in corresponding SPH measurements where the cross terms are negligible (Fig.~\ref{fig:corr_cross_lgn_sph}), we argued that this is not a convincing argument in favour of a block diagonal covariance structure, since it is not clear whether the SPH results are converged with respect to the effects of long wavelength power (see also fig.~11 of \citealp{irsic+17} which shows a mild, albeit noisy, cross-redshift correlation structure in observational data). An ideal test of correctness would involve generating skewers long enough to account for all relevant fluctuation modes as well as all possible time evolution. While this is possible with the lognormal model in principle, it is not feasible in our current implementation and we leave this as a future exercise. We would also like to underline the fact that we are focusing on an idealized scenario. In real data applications, cross-redshift bin correlations are highly impacted by the process of QSO continuum fitting. Current approaches for dealing with the unknown QSO continuum involve fitting it with a simple model from the actual data. This process results in highly correlated Ly$\alpha$ flux fluctuations along each skewer, therefore greatly affecting the small-$k$ modes of the FPS. Addressing these correlations is beyond the scope of this work.

Treating the structure predicted by the present version of the lognormal model as the truth, we then investigated the potential impact of the long wavelength modes on the inference of astrophysical parameters. We found that, although the presence of cross-redshift correlations generated by these modes in the \emph{covariance matrix} has relatively little impact, the absence of long wavelength modes in the \emph{model} can significantly impact parameter recovery, with $\gtrsim2 - \sigma$ biases seen in most of the model parameters.

Our results motivate a more detailed study of cross-redshift correlations, which we will undertake in the near future with an improved version of the lognormal model. Such a model would also enable us to perform full-fledged multi-redshift analyses \citep[e.g.,][]{Irsic_2023} for \emph{cosmological} parameter inference, robustly accounting for all relevant degeneracies with astrophysical parameters.

\section*{Acknowledgments}

We thank P. Gaikwad for help with the Sherwood simulation products. We also thank him and R. Srianand for valuable discussions. 
We gratefully acknowledge use of the IUCAA High Performance Computing (HPC) facility. We thank the Sherwood simulation team for making their data publicly available. The research of AP is supported by the Associates programme of ICTP, Trieste.

\section*{Data Availability}

The Sherwood simulations are publicly available at \url{https://www.nottingham.ac.uk/astronomy/sherwood/index.php}. The data generated during this work will be made available upon reasonable request to the authors.

\bibliographystyle{JHEP}
\bibliography{references3}

\label{lastpage}

\end{document}